\documentclass[a4paper,fleqn,usenatbib]{mnras}

\usepackage[T1]{fontenc}
\usepackage{ae,aecompl}

\usepackage{graphicx}	
\usepackage{amsmath}	
\usepackage{amssymb}	

\usepackage{color}
\usepackage{placeins} 
\usepackage{float} 
\definecolor{red}{rgb}{1,0,0}
\definecolor{black}{rgb}{0,0,0}
\definecolor{blue}{rgb}{0,0,1}
\definecolor{green}{rgb}{0,0.7,0}
\newcommand\red{\color{red}}
\newcommand\bla{\color{black}}

\title[Capture into Corotation Resonances]{Derivation of Capture Probabilities for the Corotation Eccentric Mean Motion Resonances}

\author[M. El Moutamid et al.]{
Maryame El Moutamid,$^{1,2}$\thanks{E-mail: maryame@astro.cornell.edu}
Bruno Sicardy,$^{3}$
and St\'efan Renner$^{4}$
\\
$^{1}$Cornell Center for Astrophysics, Planetary Science, Department of Astronomy, Cornell University, Ithaca, NY 14853, USA\\
$^{2}$Carl Sagan Institute, Cornell University, Ithaca, NY 14853, USA\\
$^{3}$LESIA, Observatoire de Paris, CNRS UMR 8109, Universit\'e Pierre et Marie Curie, Universit\'e Paris-Diderot,
5 place Jules Janssen,\\ F-92195 Meudon Cedex, France \\ 
$^{4}$IMCCE, Observatoire de Paris, CNRS UMR 8028, Universit\'e Lille 1, Observatoire de Lille 1 impasse de l'Observatoire, F-59000 Lille, France\\
}

\pubyear{2016}

\begin{document}
\label{firstpage}
\pagerange{\pageref{firstpage}--\pageref{lastpage}}
\maketitle

\begin{abstract}
We study in this paper the capture of a massless particle into an 
isolated, first order Corotation Eccentric Resonance (CER), 
in the framework of the Planar, Eccentric and Restricted Three-Body problem 
near a $m+1:m$ mean motion commensurability ($m$ integer).
While capture into Lindblad Eccentric Resonances (where the perturber's orbit is circular) 
has been investigated years ago, capture into CER (where the perturber's orbit is elliptic)
has not yet been investigated in detail.
Here, we derive the generic equations of motion near a CER in the 
general case where 
both the perturber and the test particle migrate.
We derive the probability of capture in that context, 
and we examine more closely two particular cases:
$(i)$ 
if only the perturber is migrating, capture is  possible only
if the migration is outward from the primary. Notably, the probability of capture is 
independent of the way the perturber migrates outward; $(ii)$
if only the test particle is migrating, 
then capture is possible only if 
the algebraic value of its migration rate is a decreasing function of orbital radius.
In this case, the probability of capture is proportional to the radial gradient of migration.
These results differ from the capture into Lindblad Eccentric Resonance (LER), 
where it is necessary
that the orbits 
of the perturber and the test particle converge for capture to be possible.
Possible applications for planetary satellites are discussed.
\end{abstract}



\section{Introduction}

Orbital captures into Mean Motion Resonance (MMR) is the key 
to understanding the orbital evolution of satellites, rings and planets. 
The special case of the Lindblad Eccentric Resonance (LER)  has been investigated 
by many authors in the context of the Planar, \textit{Circular} and Restricted Three-Body Problem.

In this case, a secondary object orbiting a massive primary body 
perturbs a massless particle, so that  the critical angle 
$\phi_L=(m+1)\lambda_s - m\lambda_p - \varpi_p$ librates, 
where $m$ is an integer, $\lambda_s$ and $\lambda_p$ are the longitudes of the secondary and the test particle, 
respectively, and $\varpi_p$ is the longitude of pericenter of the test particle.  
In a general context, \cite{henrard82}  estimated  the probability of capture 
into a first order LER, 
while \cite{borderies84} extended this work and derived 
capture probabilities into $m+1:m$ and $m+2:m$ (second order) LERs.

If the orbit of the secondary is \textit{eccentric}, 
a Corotation Eccentric Resonance (CER) appears close to and associated with each LER.
It is dynamically described by the critical angle 
$\phi_c=(m+1)\lambda_s - m\lambda_p - \varpi_s$, 
where $\varpi_s$ is the longitude of the secondary pericenter. 
The physical effects of LERs and CERs are not the same: 
the CER mainly affects the semi-major axis of the test particle and keeps its orbital eccentricity almost constant,  
forcing the particle to librate inside so-called corotation sites like a simple pendulum.
In contrast,  the LER acts on its eccentricity but keeps the semi-major axis almost constant \citep{elmoutamid14}. 

In this work, we derive the probability of capturing a test particle into an isolated CER,
as both the secondary and the particle suffer orbital migration that secularly change
their semi-major axes.
This is a novel calculation that complements what has been done before in the case of LERs.
The term ``isolated" means here that the CER and LER are sufficiently pulled apart so
that their coupling is negligible. 
%
Eventually, the goal is to extend the study of captures into MMR to more realistic cases where 
both the CER and LER act in concert on the particle, but this will not be considered here.

This work can be applied to many situations, in the context of planetary rings and satellites. 
In the Saturnian system for example, the satellites Aegaeon, Anthe and Methone are respectively  
captured into 7:6, 10:11 and 14:15 CERs with Mimas \citep{cooper08,hedman09,hedman10,elmoutamid14}. 
Atlas is in a 54:53 CER with Prometheus \citep{renner16}. 
In the case of the Neptunian system, Adams ring arcs may be dynamically confined by the satellite 
Galatea via corotation resonances \citep{renner14,nicholson95,foryta96,sicardy99}.

\if{
Moreover, thanks to the Kepler mission, it has been observed that in the context of exoplanets pairs, 
commensurabilities configurations are very rare. 
Here we show that the probability of capture is very small 
\red
[I do not see anywhere that we show that!]
\bla
which is in agreement  with  the exoplanet pairs results of \cite{fabrycky14}. 
\red
[I do not agree, see the discussion]
\bla
Future work in this context will be done to study those cases.
\red
[This whole paragraph belongs to the discussion, it is too early here].
\bla
}\fi

We note here 
that the capture of a particle into a CER bears some resemblance with
the capture of a rotating body into spin-orbit resonance \citep{gold66}. In both cases, a slow effect 
(orbital migration in our case, tidal friction for the
spin-orbit resonances) drives a pendulum-like system into
a librating state, possibly capturing it permanently into that state.
This will be commented later.

The paper is structured as follows: 
In section 2, we describe the dynamical structure of our problem based on the 
 so-called  CorALin model, taking into account the dissipation parameters.
In Section 3, we study the particle and the secondary migration terms involved in the derivation of the probability of capture.
The latter is derived in section 4. 
In section 5, we discuss the similarity of our problem with the capture in spin-orbit resonance
and finally, a summary and conclusions are given in Section 6.

\section{Dynamical structure of the problem}

We consider the restricted, planar (but \textit{not} circular) three--body problem,
in which a test particle orbits around a central mass $M_c$,
near a first order mean motion resonance $m+1:m$\footnote{The case $m$ positive (resp. negative) implies that the particle orbits inside 
(resp. outside) the secondary orbit.}
with a perturbing secondary of mass $m_s$ and an orbital eccentricity $e_s$.
The various quantities and notations used hereafter are defined in Table~\ref{tab_param_CorALin}.
\begin{table}
	\centering
	\caption{Variables, parameters and notations used in the text.}
	\label{tab_param_CorALin}
	\begin{tabular}{|*{2}{c|}}
		\hline
		Quantities  & Definitions   \\
		\hline
		$\phi_c$  & $(m+1)\lambda_s - m\lambda_p - \varpi_s$ \\ 
		$\phi_L$  & $(m+1)\lambda_s - m\lambda_p - \varpi_p$ \\
		$ h $ & $ \sqrt{3} \mid m\mid e_p \cos(\phi_L) $  \\	
		$ k $ &  $ \sqrt{3} \mid m\mid e_p \sin(\phi_L) $   \\
		$\chi$ &  $3m\Delta a/2a_0 = 3m (a_p - a_0)/2a_0$ \\
		$J_c$ & $\chi + (h^2+k^2)/2$ \\ 
		$ \epsilon_L $ & $  \sqrt{3} \mid m\mid (m_{s}/M_c) (a_0/a_s) A_m$  \\
		$ \epsilon_c $ & $ 3m^2 (m_{s}/M_c) (a_0/a_s)  A'_m e_s $  \\
		$ D $ & $ (\dot{\varpi}_s - \dot{\varpi}_p)/n_0 $  \\
		\hline
	\end{tabular}\\
\raggedright
Notes: $\lambda$, $\varpi$, $a$, $e$ denote 
the mean longitude, longitude of periapse, semi-major axis and eccentricity, respectively, and
$\dot{\varpi}$ denotes the secular apsidal precession rate
(e.g. forced by the oblateness of the central body or other perturbing secondaries).
Subscripts $s$ and $p$ refer to the secondary and test particle, respectively.
Subscript 0 is used for quantities estimated at the exact corotation radius, $a_0$.
The quantities $A_m$ and $A'_m$ are combinations of Laplace coefficients
with $A_m  \sim -A'_m \sim 0.8m$ for large $|m|$'s.
The quantities $\dot{\varpi}$ and $\dot{\varpi_s}$ are the rate of precession of the orbits of the test particle and the secondary, respectively. 
\end{table}

Near the resonance, the equations of motion reduce to a two degree of freedom system 
associated with the two critical angles 
$\phi_c$  and $\phi_L$.
The subscripts \textit{c} and \textit{L} refer to eccentric corotation and Lindblad resonances
(CER and LER respectively).
Details are given by \cite{elmoutamid14}, 
who encapsulated the equations of motion of the test particle 
in the so-called CorALin model%
\footnote{The presentation of the CorALin model here is slightly different from the one given in 
\cite{elmoutamid14}. The main differences are the sign of $\phi_L$ and the presence 
of the parameter $\epsilon_c$, which provides a better understanding of the physical effect
of a corotation resonance.}: %
\begin{equation}
\left\{
\begin{array}{ll}
dJ_c/d\tau      =  &  -\epsilon_c\sin(\phi_{c}) \\
d\phi_c/d\tau = & \chi  \\
dh/d\tau          = & -(\chi + D)  k  \\
dk/d\tau          = & +(\chi + D) h  + \epsilon_L, \\
\end{array}
\right.
\label{eq_allchi}
\end{equation}
where $\tau= n_0 t$ is a dimentionless time scale, and $n_0$ is the mean motion
at exact corotation\footnote{Note that the time scale used later in this paper is the usual time $t$,  and not $\tau$.}.
The strengths of the corotation and Lindblad resonances are quantified by the parameters
$\epsilon_c$ and $\epsilon_L$, respectively, see Table~\ref{tab_param_CorALin}.

The CorALin model portrays two coupled resonances: 
a CER described by a simple pendulum system (first two equations in Eqs.~(\ref{eq_allchi})), 
coupled to the  LER given by the last two equations 
(the so-called second fundamental model for resonance). 
The dimensionless parameter $D$ measures the distance (\textit{in frequency}) 
between the two resonances and acts as a coupling parameter between the two.
For large $D$'s, the resonances decouple and for $\chi=0$, 
only the simple-pendulum motion is relevant, 
while the particle eccentricity $e_p \propto \sqrt(h^2+k^2)$ remains constant.

\begin{center}
\begin{figure*}
\includegraphics[width=17cm,height=8cm]{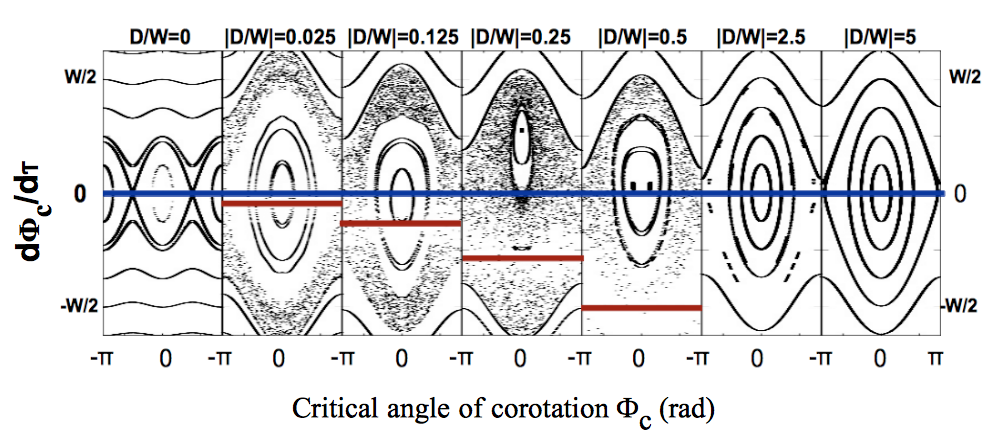}
 \caption{Poincar\'e surfaces of section of system (\ref{eq_allchi}) using $\epsilon_L=-0.1$, $\epsilon_c=1$ and $m=1$ 
 (resonance 2:1) for seven different values of $D$. Each section is obtained when $k=0$ and $\dot{k}>0$ 
 for 15 different initial conditions with the same value of energy. Note that $\tau=n_0t$ is a dimensionless time scale (more details are given in the text). For $D=0$ and large $D$'s,  
 the trajectories are regular, while chaos is dominant for intermediate cases, see more details in \citealt{elmoutamid14}.
The blue lines represent the position of the center of CER, while the red lines represent the position of the center of the LER.
 }
 \label{fig_chaos}
\end{figure*}
\end{center}

Stable oscillations of $\phi_c$ occur around $\phi_c = 0$ (resp. $\phi_c = \pi$) for $\epsilon_c$ 
positive (resp. negative), 
with periods $\sim 2\pi/\sqrt{|\epsilon_c|}$ in the CorALin time unit, or 
$\sim 2\pi/(n_0\sqrt{|\epsilon_c|})$ in the usual time unit. 
Moreover, the full width (in units of $\chi$) of the corotation site is $W= 4\sqrt{|\epsilon_c|}$, or 
\begin{equation}
W_{\rm CER} = 
\frac{8a_0\sqrt{|\epsilon_c|}}{3|m|}
\label{eq_W}
\end{equation}
in physical distance units\footnote{Note that $m$ and $\epsilon_c$ have opposite signs.}.

The decoupling between the CER and LER occurs for 
\begin{equation}
D_{\rm CL} > \sim W_{\rm CER},
\label{eq_condition}
\end{equation}
where $D_{\rm CL}$ is the radial splitting
between the resonances and given by:
\begin{equation}
D_{\rm CL} = 
\frac{2a_0}{3m}\left(\frac{\dot{\varpi}_s - \dot{\varpi}_p}{n_0}\right),
\label{eq_D}
\end{equation}
The condition~(\ref{eq_condition}) is fulfilled in the right-most panels of  Fig.~(\ref{fig_chaos}).
That figure shows that for intermediate values of $D_{\rm CL}$, 
the CER and LER are superimposed and strongly coupled, leading to chaotic behavior.  
However, the system is again an integrable one as $D_{\rm CL} =0$.

We will restrict our study to the case of large $D_{\rm CL}$'s only (Eq.~\ref{eq_condition}),
and consider the effect of orbital migrations of \it both \rm the secondary and the test particle.
This causes a secular variation of $\chi$ (beyond the effect of the CER)
since both $a_p$ and $a_0$  slowly change.

At this point, it is important to note that $\chi$ 
(a local form of the particle semi-major axis, see Table~\ref{tab_param_CorALin})
is \textit{not} the appropriate action variable to use together with the angle $\phi_{c}$. 
Instead, an angular momentum-type quantity should be used to avoid 
cumbersome additional terms in the equations of motion in the presence of migration,
and permit the use of adiabatic invariance arguments.
Here we choose the variable $J_\phi= 3m(J_p-J_0)$ as the conjugate of $\phi_c$,
where $J_p$ is the specific angular momentum of the particle and
$J_0=\sqrt{G M_c a_0}$ is the specific angular momentum at radius $a_0$
($G$ being the gravitational constant).
As the CER has a very small effect on eccentricity in the absence of LER \citep{elmoutamid14},
we can assume here that the particle has a circular orbit, then $J_{\phi}= \chi J_0$ is proportional to $\chi$.

\begin{figure*}
		\includegraphics[width=5.84cm,height=7cm]{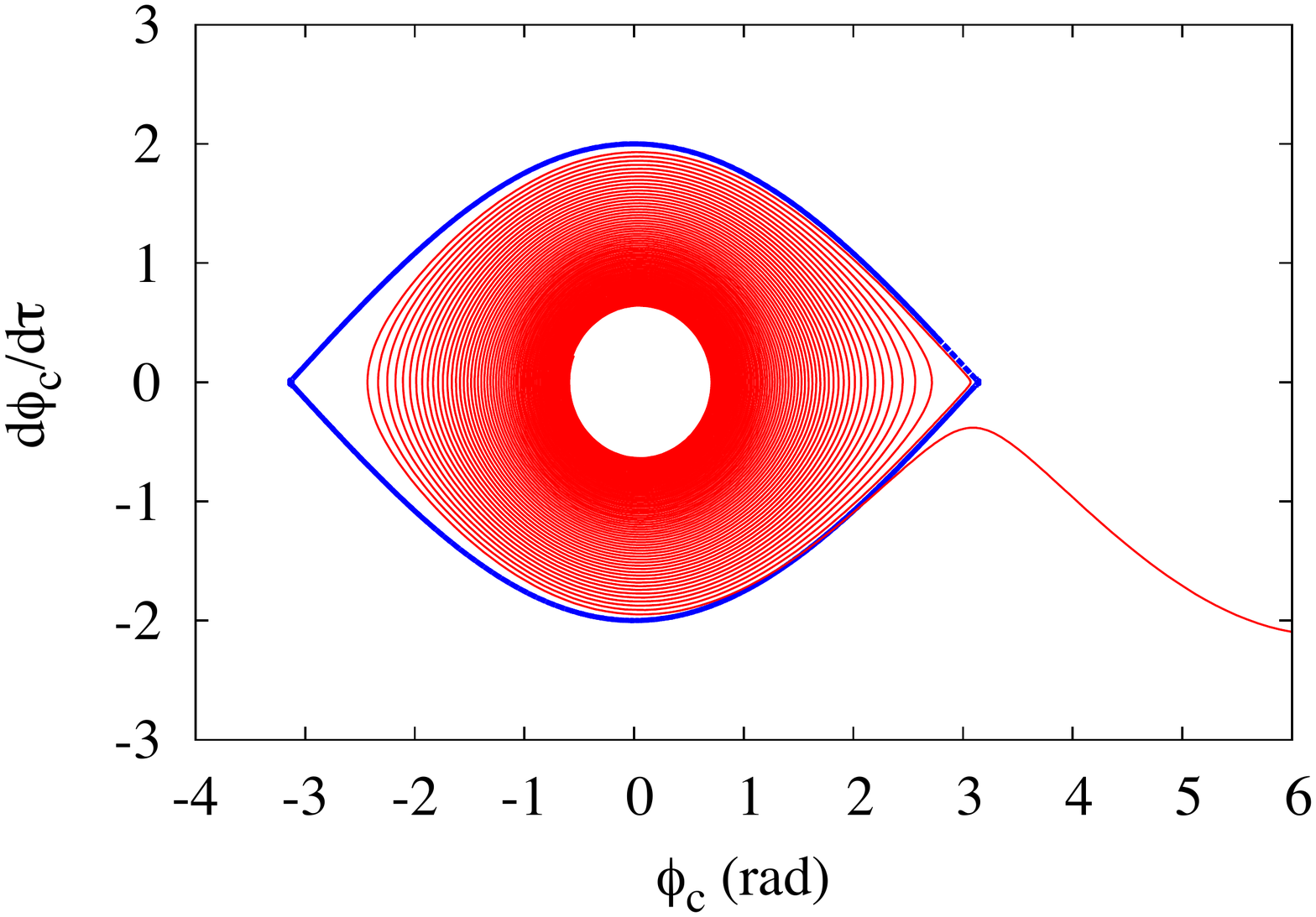}
		\includegraphics[width=5.84cm,height=7cm]{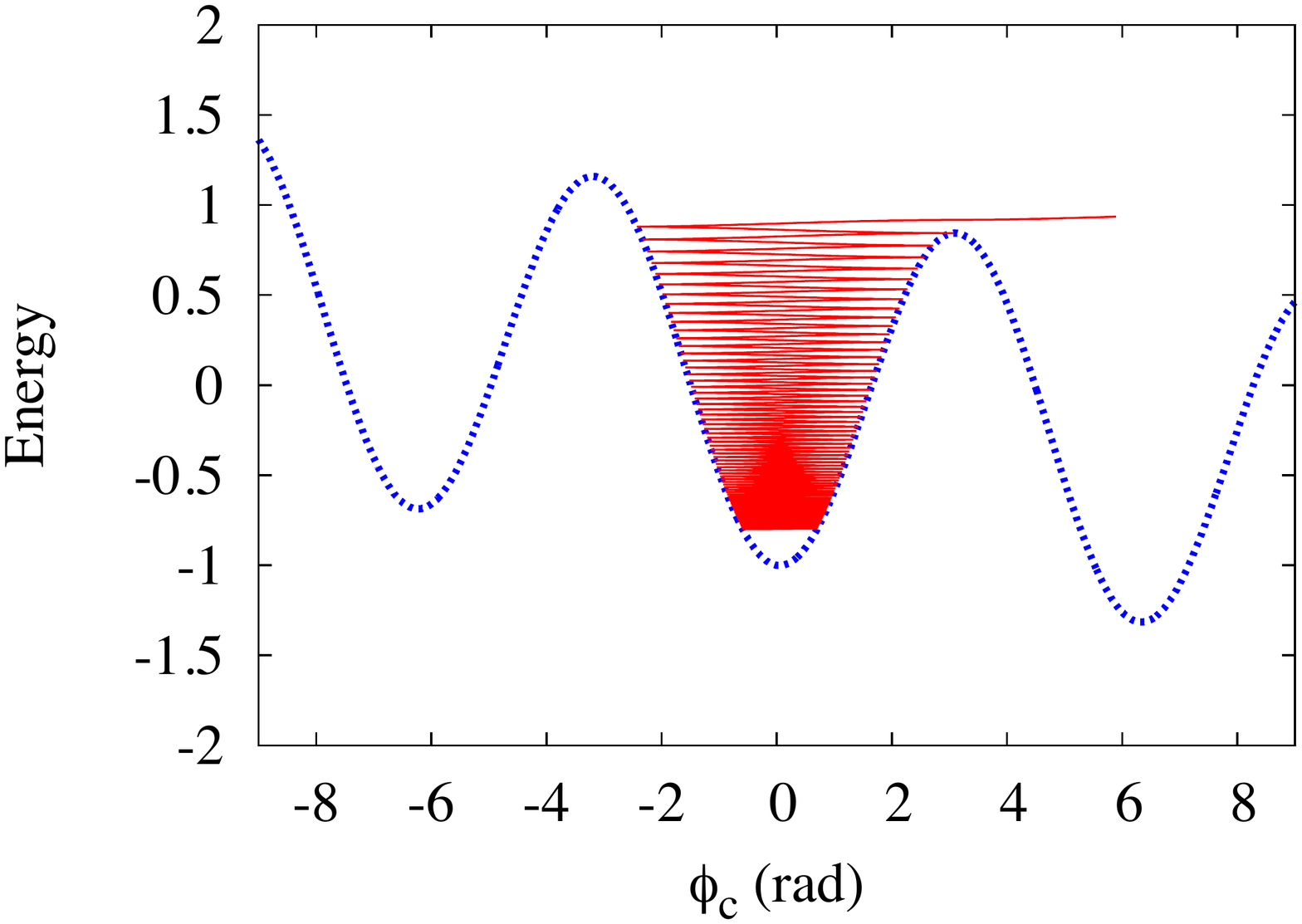}
		\includegraphics[width=5.84cm,height=7cm]{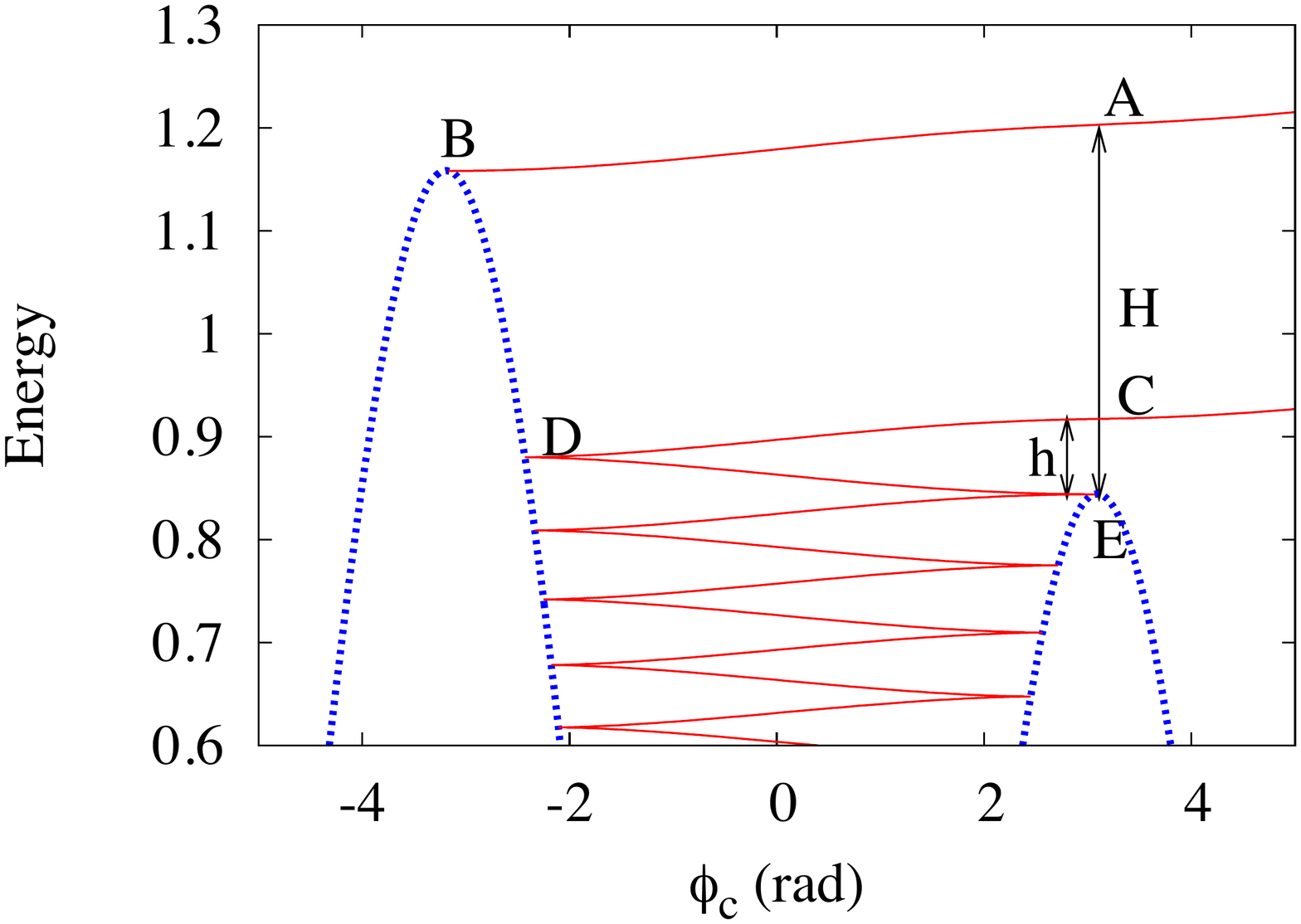}
    \caption{An example of a successful capture in CER, considering a case where $\epsilon_c > 0$.
    The parameters ($\epsilon_c=1$, $\epsilon_{mig}=5\times10^{-2}$ and $d{\epsilon}_{mig}/d\chi=-5\times10^{-3}$) 
    have been greatly exaggerated (compared to practical cases) for better viewing. 
    Note that $\tau=n_0t$ is a dimensionless time scale (more details are given in the text). 
    \textbf{Left panel:} Phase space $(\phi_c,\dot{\phi}_c)$, where the blue curve represents the separatrix, 
    and the red curve is the captured trajectory illustrated in the right panels.
    \textbf{Middle panel-} Blue curve: the potential energy $U$ (Eq.~\ref{eq_pot}) versus $\phi_c$.
    Red curve: the total energy $E$ (Eq.~\ref{eq_E}).
    \textbf{Right}: vertical expansion of the middle panel, showing the parameters $h$ and $H$ used for the
    calculation of the probability of capture, see text.
    }%
    \label{fig_capture}
\end{figure*}

In the absence of migration, 
the Hamiltonian that locally describes the simple-pendulum motion near the CER reads:
\begin{equation}
{\cal H}_0 = \left(\frac{GM_c}{J_0}\right)^2 
\left[ 
\frac{1}{2} \left(\frac{J_\phi}{J_0}\right)^2
- \epsilon_c \cos(\phi_c)
\right],
\label{eq_H1}
\end{equation}
with the correct dimension of a specific energy (see \cite{sicardy03} for details on the method).

We now define $\epsilon_s$ (resp. $\epsilon_p$) 
as the dimensionless secular migration rates of the 
secondary (resp. test particle)\footnote{%
Strictly speaking, $\epsilon_s$ actually measures the migration rate of the resonance
radius $a_0$, related to the secondary migration rate by $da_0/dt = [m/(m+1)]^{2/3}(da/dt)$.}%
: 
\begin{equation}
\begin{array}{l}
\displaystyle
 \epsilon_s= \frac{1}{n_0 a_0} \cdot \frac{da_0}{dt} = \frac{2}{n_0J_0}  \left (\frac{dJ_0}{dt}\right)_{\rm mig} \\
 \\
\displaystyle
 \epsilon_p= \frac{1}{n_0 a_0} \cdot \frac{da_p}{dt} = \frac{2}{n_0J_0}  \left (\frac{dJ_p}{dt}\right)_{\rm mig}, \\
\end{array}
\end{equation}
where the ``mig'' index means the secular effect of the migration on the variation of $J_0$ and $J_\phi$.

Those migration rates introduce additional terms
$\dot{J}_\phi = (3m/2)n_0 J_0 (\epsilon_p - \epsilon_s)$ in the equations of 
motion\footnote{Note that $\epsilon_c$ does \textit{not} vary during the migration of the secondary,
due its mere definition (Table~\ref{tab_param_CorALin}).}.
They can be incorporated in ${\cal H}_0$ to form a new Hamiltonian:
\begin{equation}
{\cal H} = \left(\frac{GM_c}{J_0}\right)^2 
\left[ 
\frac{1}{2} \left(\frac{J_\phi}{J_0}\right)^2
- \epsilon_c \cos(\phi_c)
+\frac{3}{2}m\epsilon_{\rm mig}\phi_c
\right],
\label{eq_H2}
\end{equation}
where
\begin{equation}
\epsilon_{\rm mig} = \epsilon_s-\epsilon_p.
\label{eq_epsilon_mig}
\end{equation}

Note the mirror-symmetry between the factors $\epsilon_s$ and $\epsilon_p$:
as expected, the migration of the secondary in one direction has the same effect as 
the migration of the particle in the opposite direction. 
Also, the Hamiltonian nature of the motion 
means that the particle \it cannot \rm be captured in the CER: 
$\phi_c$ is either permanently librating 
or permanently circulating.
This has a simple physical interpretation: the particle approaches the corotation
radius at the same rate as it recedes from it. This symmetry prevents capture.

However, as $\epsilon_s$ is in general \it time\rm-dependent \rm
and the particle migration is usually \it space\rm-dependent, capture becomes possible.
Here we make the simplest assumption that the particle migration has a local gradient 
parametrized by a dimensionless parameter $\epsilon_g$, that we define as follows:
\begin{equation}
\frac{1}{n_0 a_0} \frac{da_p}{dt}= 
\epsilon_p + \epsilon_g \left( \frac{a_p - a_0}{a_0}\right).
\label{eq_grad}
\end{equation}
As we see next, the factors $\epsilon_s$ and $\epsilon_g$ break the symmetry 
between the approach to and the recession from CER, possibly leading to capture.
In the rest of the paper, all the terms involving the letter $\epsilon$ will be
assumed to be small compared  to unity, corresponding to the fact that the
migration rates and their variations are assumed to be small.

\section{Migration}

The shape of the particle trajectory in the ($J_\phi, \phi$) phase space, 
in particular the fact that it is in libration or circulation, solely depends
on the value of the term into brackets in Eq.~(\ref{eq_H2}):
\begin{equation}
\begin{array}{ll}
E & 
\displaystyle 
=
\frac{1}{2} \left[\frac{J_\phi}{J_0(t)}\right]^2 
-\epsilon_c\cos(\phi_c)
+\frac{3}{2}m\epsilon_{\rm mig}(t)\phi_c \\
& \\
& 
\displaystyle 
=
\frac{1}{2n_0^2} \dot{\phi}_c^2 
-\epsilon_c\cos(\phi_c)
+\frac{3}{2}m\epsilon_{\rm mig}\phi_c \\
\end{array}
\label{eq_E}
\end{equation}
We have enhanced in the first equation that  both $J_0$ and $\epsilon_{\rm mig}$
explicitly depends upon time, due to migration, and
we have used $\dot{\phi}_c = n_0(J_\phi/J_0)$ to write the second equation.

The quantity $E$ can be viewed as the dimensionless ``energy" of a 
particle moving into the ``potential" $U$ of a modified simple pendulum, where:
\begin{equation}
U= -\epsilon_c\cos(\phi_c) + \frac{3}{2}m \epsilon_{\rm mig}\phi_c.
\label{eq_pot}
\end{equation}
The function $U(\phi_c)$ is plotted in blue in the two right-most 
panels of Fig.~\ref{fig_capture}.

The two slowly varying parameters, $J_0$ and $\epsilon_{\rm mig}$, 
together with the term $\epsilon_g$, cause a variation of the energy $E$: 
\begin{equation}
\dot{E}= \frac{dE}{dt}= 
-\frac{ \epsilon n_0}{2} \left(\frac{J_\phi}{J_0}\right)^2 + \frac{3}{2} m \dot{\epsilon}_{\rm mig} \phi_c =
-\frac{\epsilon}{2n_0}  \dot{\phi}_c^2 + \frac{3}{2} m \dot{\epsilon}_{\rm mig} \phi_c,
\label{eq_dEdt}
\end{equation}
where $\epsilon= \epsilon_s - 2 \epsilon_g$.

A necessary condition of capture is that the particle loses energy $E$, in order
to converge toward a local minimum of $U$.
Ignoring the term in $\dot{\epsilon}_{\rm mig}$ 
(see Appendix), this requires: 
\begin{equation}
\epsilon= \epsilon_s - 2\epsilon_g > 0.
\label{eq_epsilon}
\end{equation}

A paradoxical result is that $\epsilon$ depends on the 
\it migration rate \rm $\epsilon_s$  of the secondary and on the
\it gradient \rm $\epsilon_g$ of the particle migration rate. 
In other words, there is a symmetry breaking between the respective 
effects of the secondary and the particle migrations.
This stems from the fact that we consider here the restricted problem.

This effect is described \cite{sicardy03}, who considered  the different, 
but connected problem of two co-orbital secondary masses $m_1$ and $m_2$ suffering 
slow migrations with rates $\epsilon_1$ and $\epsilon_2$, and gradients
$G'_1$ and $G'_2$ (equivalent to the term $\epsilon_g$ used here), respectively.
Their equation A.13 shows that $\dot{E}$ stems from two terms in front of $\dot{\phi}_c^2$, 
one proportional to $(m_1 \epsilon_1 + m_2 \epsilon_2)/(m_1+m_2)$ and
one proportional to $(m_1 \epsilon_2  G'_2 + m_2 \epsilon_1  G'_1)/(m_1+m_2)$.
As long as $m_1$ and $m_2$ are non-zero, there is a commutativity between 
the effects of the migrations of the two masses.
As $m_1$, say, tends to zero, however one obtains a term in $\epsilon_2$ only 
(equivalent to the term in $\epsilon_s$ in Eq.~\ref{eq_epsilon}) and 
a term in $\epsilon_1  G'_1$ only (equivalent to the term in $\epsilon_g$).

Turning back to  Eq.~(\ref{eq_epsilon}), we see that if acting alone ($\epsilon_g=0$),  
the secondary migration can lead to capture only \it if the migration is outwards \rm ($\epsilon_s > 0$).
If the particle migration is acting alone ($\epsilon_s=0$), 
capture is possible only \it if its gradient is negative \rm ($\epsilon_g < 0$)\footnote{%
Note that this gradient concerns the \textit{algebraic} value of the migration rate, 
not its absolute value. For instance an inward migration rate ($\epsilon_p < 0$) whose
absolute value decreases with orbital radius will have a positive gradient  ($\epsilon_g > 0$).
}. %
In that case, the particle approaches the CER faster than it recedes from it, permitting the capture.

Once the capture is effective, the particle will converge towards a local minimum of $U$, 
i.e. the CER center (Fig.~\ref{fig_capture}).
Considering the effect of secondary migration only, one can use an adiabatic invariant argument to
estimate how the particle converges to that minimum. 
Using for instance $\epsilon_c > 0$ and considering a motion close to the stable point  $\phi_c=0$,
we have $\cos(\phi_c) \sim 1-\phi_c^2/2$, and thus a harmonic motion of the type
$J_\phi = {\bf  J_\phi}  \cos(\omega t)$ and 
$\phi_c = {\bf \Phi_c}  \sin(\omega t)$, 
with $\omega= n_0 \sqrt{\epsilon_c}$ and
where boldfaces denote amplitudes.
The equations 
$\dot{J}_\phi = -\partial {\cal H}/\partial \phi_c$ and 
$\dot{\phi}_c = +\partial {\cal H}/\partial J_\phi$ then provide
${\bf  J_\phi} =  \sqrt{\epsilon_c} J_0 {\bf \Phi_c}$,
so that the action is ${\cal A}= {\bf J_\phi \boldsymbol \Phi_c} = \sqrt{\epsilon_c} J_0  {\bf \Phi^2_c}$,
where $J_0 = \sqrt{GM_c a_0}$.
For a slow variation of $a_0$, ${\cal A}$ is adiabatically conserved so that 
$\sqrt{a_0}  {\bf  \Phi^2_c}$= constant, thus:
\begin{equation}
{\bf \Phi_c} \propto \frac{1}{a^{1/4}_0},
\label{eq_adiabatic}
\end{equation}
a result already obtained in the case of co-orbitals bodies \bla by \cite{fleming00} and \cite{sicardy03}.
Note that the convergence toward the CER center thus weakly depends upon $a_0$.

If the particle migration is acting alone, one obtains easily from the previous calculations that
$d{\bf \Phi_c}/dt= (n_0\epsilon_g/2){\bf \Phi_c}$, so that 
\begin{equation}
{\bf \Phi_c} \propto \exp(n_0 \epsilon_g t/2),
\label{eq_convergence}
\end{equation}
where we recall that $\epsilon_g$ must be negative for capture to occur. 
Then the particle converges towards the CER center on a time scale 
$t \sim 2/(n_0 |\epsilon_g|)$. 

When both the secondary and the particle migrate, the two equations above 
can be combined to provide:
\begin{equation}
{\bf \Phi_c} \propto \frac{\exp(\int\! \epsilon_g n_0 dt/2)}{a^{1/4}_0}.
\label{eq_combined}
\end{equation}

Note that the evaluations above can be used to estimate on which time scales
the particle \it escapes \rm the CER libration center for the cases $\epsilon < 0$.
This may happen if the particle is formed inside the CER site, e.g. a dust particle
ejected from the surface of a body currently trapped in a CER.

\section{Probability of capture}

Fig.~\ref{fig_capture} illustrates the capture mechanism.
It closely follows the path described in \cite{gold66} and \cite{murray99} for capture
in spin-orbit resonances.
The right-most panel shows that the particle must approach the corotation 
site through a small energy ``window" $h$, inside the possible interval $H$.
The particle moves in that plot along the red curve with slope:
\begin{equation}
\frac{dE}{d\phi_c}= \frac{1}{\dot{\phi}_c}\frac{dE}{dt} = 
-\frac{\epsilon}{2n_0}\dot{\phi}_c,
\label{eq_dedphic}
\end{equation}
where we have neglected the term in $\dot{\epsilon}_{\rm mig}$ in Eq.~(\ref{eq_dEdt}), 
see Appendix.

The trajectories AB and CDE bound the extreme trajectories that will escape the CER.
As those trajectories come very near the hyperbolic points at $\phi_c \approx \pi$ and $-\pi$, they have
energy $E \approx +\epsilon_c$, so that
$\dot{\phi}_c\approx \pm n_0 \sqrt{2|\epsilon_c |(1+\cos(\phi_c))} = \pm 2n_0 \sqrt{|\epsilon_c|}\cos(\phi_c/2)$.
Injecting that expression into Eq.~(\ref{eq_dedphic}), 
one can integrate $dE/d\phi_c$ to obtain the variations of $E$ alongs the trajectories AB and CDE.
Elementary calculations based on the form of $U$ (Eq.~\ref{eq_pot})
then provide:
\begin{equation}
h= 8\epsilon \sqrt{|\epsilon_c|} {\rm ~~and~~}
H= 3\pi |m \epsilon_{\rm mig}| + 4\epsilon \sqrt{|\epsilon_c|}, \\
\label{eq_hH}
\end{equation}
from which the probability of capture $P_m= h/H$ into the $m+1:m$ CER is derived,
assuming that the initial energy of the particle is randomly and uniformly distributed in the ``window" $H$.
Substituting the value of $\epsilon_c$ as given in Eq.~(\ref{eq_W}) we obtain:
\begin{equation}
P_m=\frac{h}{H}= 
\frac{2\epsilon W_{\rm CER}}{2\pi a_0 |\epsilon_{\rm mig}| + \epsilon W_{\rm CER}}, 
\label{eq_proba_gen}
\end{equation}
where we recall that
$\epsilon = \epsilon_s - 2\epsilon_g$ and $\epsilon_{\rm mig}= \epsilon_s - \epsilon_p$.
We also recall that the condition $\epsilon > 0$ must be fulfilled for the capture to occur 
(Eq.~\ref{eq_epsilon}).

At this point, we note that in general $W_{\rm CER} \ll a_0$,
since $\epsilon_c$ is very small in Eq.~\ref{eq_W}, 
as $m_s/M_c$ and $e_s$ are themselves usually small (Table~\ref{tab_param_CorALin}).
Moreover, in general, the particle migration rate $da_p/dt$ 
does not change rapidly with distance. 
Taking for instance $da_p/dt = K/a_p^q$, where $K$ is a constant and $q$ 
 is usually of order unity,
we obtain from Eq.~(\ref{eq_grad}) $\epsilon_g= -q \epsilon_p$.
As a consequence $\epsilon_g$ and $\epsilon_p$ are usually of same order,
so that the term in $W_{\rm CER}$ in the denominator of the equation above can be neglected.
Thus, the probability of capture when both the secondary and the particle migrate is:
\begin{equation}
P_m \approx
\frac{W_{\rm CER}}{\pi a_0} \left| \frac{\epsilon}{\epsilon_{\rm mig}}\right| =
\frac{W_{\rm CER}}{\pi a_0} \left| \frac{\epsilon_s-2\epsilon_g}{\epsilon_s-\epsilon_p}\right|.
\label{eq_proba_gen_approx}
\end{equation}

We now examine the two extreme cases where 
$(i)$ only the particle migrates and where
$(ii)$ only the secondary migrates.
In case $(i)$, we have $\epsilon_s=0$, so that
\begin{equation}
\begin{array}{ll}
\displaystyle
P_m \approx 
\frac{2W_{\rm CER}}{\pi a_0} \left| \frac{\epsilon_g}{\epsilon_p}\right| = & \\
& \\
\displaystyle 
\frac{2W_{\rm CER}}{\pi a_{\epsilon p}} & {\rm (particle~migration~only)} \\
\label{eq_proba_eg}
\end{array}
\end{equation}
where we define $a_{\epsilon p}= a_0 |\epsilon_p/\epsilon_g|$ as
the radial scale over which the particle migration rate suffers significant changes.
We recall here that the condition $\epsilon_g < 0$ must be fulfilled. 

In case $(ii)$, we have $\epsilon_p= \epsilon_g= 0$ (and $\epsilon_s > 0$), so that
\begin{equation}
\begin{array}{ll}
\displaystyle
P_m \approx \frac{W_{\rm CER}}{\pi a_0} & 
{\rm (secondary~migration~only)} \\
\label{eq_proba_es}
\end{array}
\end{equation}
with the noteworthy result that in this case the probability of capture is
\it independent of the way the secondary migrates. \rm
In other words, we do not need to know the details
of that migration to derive the probability of capture.
Note also from Eq.~(\ref{eq_W}) and Eq.~(\ref{eq_proba_es}) that $P_m$ is \it independent of $a_0$, \rm
since $\epsilon_c$ is a constant.
In fact, $P_m$ now only depends on $\epsilon_c$ and $m$
(i.e. the secondary mass and orbital eccentricity, plus some geometrical factors, see Table~\ref{tab_param_CorALin}).

Eq.~(\ref{eq_proba_es}) can be re-written by 
noting that the physical area enclosed in each corotation site (see the physical area enclosed 
in the blue separatrix in the left-most panel of Fig.~\ref{fig_capture}) is 
$4 a_0 W_{\rm CER}/|m|$. 
Since there are $|m|$ such corotation sites,
the total area occupied by the CER is $S_{\rm CER} = 4 a_0 W_{\rm CER}$.
Moreover, the area enclosed of an orbit of semi-major axis $a_0$ is $S_0=\pi a_0^2$, so that:
\begin{equation}
P_m \approx \frac{S_{\rm CER}}{4S_0}.
\label{eq_proba_SCER}
\end{equation}

Thus, if only the secondary migrates, the probability of capture is one quarter
of the area occupied by the CER libration sites divided by the area enclosed in the CER orbit.


\section{Similarities with spin-orbit resonances}

At this stage, 
we note that the capture of a particle into a CER resembles the capture of a rotating body into spin-orbit resonance 
(see \citealt{gold66} or \citealt{murray99}). 
This follows from the fact that the dynamical structures of CERs and spin-orbit resonances are both based 
on simple pendulum systems. 
The capture happens because of orbital migration in the case of a CER, 
and tidal friction (and thus migration in the frequency space) for spin-orbit resonances. 
Using Eq.~\ref{eq_W} and the fact that the libration frequency around the corotation center is 
$\omega_0= n_0 \sqrt{|\epsilon_c|}$, one can re-write eq.~\ref{eq_proba_gen} as:
\begin{equation}
P_m=
\frac{4(\omega_0/n_0)}{(3/4) \pi a_0 |m| |\epsilon_{\rm mig}/\epsilon| + 2(\omega_0/n_0)}.
\label{eq_our_proba}
\end{equation}
\cite{murray99} provide the probability of capture into a spin-orbit resonance
$\Omega = p n_0$ (where  $\Omega$ is the spin rate of the body and $p$ is a rational)
in two different cases. 
One of them  
corresponds to a tidal torque that is dependent on the frequency of each tidal component 
(the frequency in this context means the one with which the body experiences the tidal distortion),
$P_p =  4(\omega_0/n_0)/[\pi V + 2(\omega_0/n_0)]$, see  their  equation~(5.118).
The factor $V$ enters in expression of the average, frequency-dependent tidal torque acting on the body
$\langle N_s \rangle = -K (V + \dot{\gamma}/n_0)$, where $K$ is a constant
and $\gamma = \theta - pM$ is the spin-orbit resonant angle that relates the 
body orientation in inertial space, $\theta$, and its mean anomaly, $M$.
By writing the tidal torque
$\langle N_s \rangle = N_{s0} + N_{sg} (\dot{\gamma}/n_0)$, where 
$N_{s0}$ is the local tidal torque and $N_{sg}$ is its local gradient (in the frequency space),
we obtain $V= N_{s0}/N_{sg}$. Therefore, $P_p$ can be written:
\begin{equation}
P_p=
\frac{4(\omega_0/n)}{\pi (N_{s0}/N_{sg}) + 2(\omega_0/n)}, 
\label{eq_spin_proba}
\end{equation}
identical to Eq.~\ref{eq_our_proba} after posing 
$N_{s0}/N_{sg} = (3|m|/2) |\epsilon_{\rm mig}/\epsilon|$.
In the special case where $\epsilon_s= 0$ (i.e. when only the particle migrates), 
we obtain $N_{s0}/N_{sg} = (3|m|/4) |\epsilon_p/\epsilon_g|$,
which enhances the essential similarities between the CER and
spin-orbit captures.
Both describe  a particle that migrates  along the frequency axis 
of the simple-pendulum phase space at rate $N_{s0}$, with a 
local gradient $N_{sg}$ along that axis.


\section{Discussions and Conclusions}

We have studied the mechanism of capture into an isolated first order Corotation Eccentric Resonance (CER) 
in the context of the Elliptical Restricted Three-body Problem, in which a test particle orbits around a central mass $M_c$,
near a first order mean motion resonance $m+1:m$ with a perturbing secondary of mass $m_s$. 
We derive a formula for the probability of capture given by Eq.~\ref{eq_proba_gen_approx} in a general context. 
Then we apply this result to two particular cases, 
first where we consider a migration of the particle only (Eq.~\ref{eq_proba_eg}), and 
secondly, in the case of migration of the secondary (Eq.~\ref{eq_proba_es}). 
We point out the noteworthy fact that in this case,  
the capture probability does not depend on the way the secondary migrates. 

Under realistic assumptions, we rewrite our formula for capture in the second case
as Eq.~(\ref{eq_proba_SCER}). 
This equation has the interesting consequence that if the corotation radius $a_0$ 
sweeps a total area $S$ in a region where particles are uniformly distributed 
with surface density $N$, then the corotation sites will eventually be populated (due to captures)
with a particle surface density of $NS/4S_0$. Conversely, if we observe today a certain number of particles trapped in 
corotation sites, we may estimate the number of particles originally present in that region. 

This work can be applied in many cases, specially in the context of planetary rings, satellites and in the context of exoplanets.
As an example, in the context of the Saturn system, \cite{elmoutamid14}
study the case of Aegaeon which is trapped in a 7:6 CER caused by Mimas, 
with $a_0 \sim 167,500$~km. Since $W_{\rm CER} \sim 30$~km, we find a probability of capture of $\sim5\times10^{-5}$. 
This very low probability suggests that originally, there were many more such objects in the Saturn system, 
but that only a few of them were captured into CERs. 

Note finally that this study assumes that the CER is isolated, i.e. that the condition~(\ref{eq_condition})
is fulfilled. 
In the cases where $D_{\rm CL} \sim W_{\rm CER}$, the motion in phase space is chaotic, 
as  the LER and CER are strongly coupled, see Fig.~\ref{fig_chaos}. 
This occurs when the effects of the oblateness of the central body, 
the presence of a massive disk or another companion in the system, 
are not strong enough to split and thus isolate the resonances from each other. 
The chaotic nature of motion then prevents an easy analytical derivation
of the probability of capture into the $m+1:m$ mean motion resonance.
In that case, numerical integrations might be useful to see whether the 
probability derived here (Eq.~\ref{eq_proba_gen})  remains valid,
at least in order of magnitude.

\section*{Acknowledgements}

The authors are thankful to Philip D. Nicholson, Aur\'elien Crida, Damya Souami and Matthew M. Hedman for many interesting discussions on this topic.
They thank the Encelade working group for interesting discussions and the International Space Science Institute (ISSI) for support.
This work was supported by NASA through the Cassini project. 
Part of the research leading to these results has received funding from the European Research Council under the European Community's H2020 (2014-2020/ ERC Grant Agreement n\textsuperscript{o}~669416 ``LUCKY STAR'').




\appendix

\section{Effect of the variation of migration rate with time}

If the secondary migrates, and since that migration usually depends on the parameter
$a_0$,  $\epsilon_s$ also depends on time, as well as $\epsilon_{mig}= \epsilon_s - \epsilon_p$. 
In that case, the term 
$(3m/2)\dot{\epsilon}_{mig} \phi_c=
(3m/2)\dot{\epsilon}_s \phi_c$ 
in Eq.~(\ref{eq_dEdt}) causes a variation of $E$ that should be compared to 
$-\epsilon \dot{\phi}_c^2/2n_0 = -2 n_0 \epsilon |\epsilon_c| \cos^2(\phi_c/2)$.
In order to simplify the analysis, we assume a simple functional form $da_0/dt= K/a_0^q$, 
as is done after Eq.~(\ref{eq_proba_gen}).
Thus, $\dot{\epsilon}_s= -(q-1/2)n_0 \epsilon_s^2$.

Consequently, in order to neglect
$(3m/2)\dot{\epsilon}_m \phi_c$ with respect to 
$2 n_0 \epsilon |\epsilon_c| \cos^2(\phi_c/2)$,
we must have $|m|(q-1/2)(\epsilon_s/\epsilon_c) \ll 1$.
Defining the orbital migration time scale as $t_{\rm mig}= a_0/\dot{a}_0$,
denoting $T_0$ the orbital period ($2\pi/n_0$), and
using Eq.~(\ref{eq_W}), the condition above reads
\begin{equation}
\frac{q-1/2}{|m|} \left( \frac{a_0}{W_{\rm CER}} \right)^2 T_0 \ll t_m.
\label{eq_tmig}
\end{equation}

Equivalently, this requires that the radial velocity $v_{\rm mig}= da_0/dt$ 
resulting from the migration of the secondary fulfills the condition:
\begin{equation}
v_{\rm mig} \ll 
\frac{|m|}{2\pi(q-1/2)} \left( \frac{W_{\rm CER}}{a_0} \right)^2 v_0,
\label{eq_vmig}
\end{equation}
where $v_0= 2\pi a_0/T_0$ is the orbital velocity.

Those conditions must be checked on a case-by-case basis.
For instance, \cite{elmoutamid14}
study the case of Aegaeon which is trapped in a 7:6 CER caused by Mimas, with 
$m=6$, $a_0 \sim 167,500$~km, $W_{\rm CER} \sim 50$~km and $T_0 \sim 0.81$~days.
If we consider a Mimas migration caused by tides raised on Saturn, 
then $q= 11/2$ \citep{murray99}.
The condition~(\ref{eq_tmig}) thus requires $t_m \gg 20,000$~years. 
This is satisfied by a safe margin, as typical migration time scales for Mimas are 
of the order of several millions years \citep{charnoz11,crida12}.



\bsp	
\label{lastpage}
\end{document}